\documentclass[12pt]{iopart}
\usepackage{epsfig}
\usepackage{amssymb}
\newcommand{\R}{\mbox{$\mathbb R$}}
\newcommand{\adj}{^{\dag}}
\newcommand{\xreg}{\mbox{$x_{\mbox{reg}}$}}
\newcommand{\Xnew}{\mbox{$X'$}\mskip -3mu}
\newcommand{\Cnew}{\mbox{$C'$}\mskip -3mu}

\begin{document}
\jl{5}
\title[Best estimate of power]{Is the best estimate of power equal to\\
the power of the best estimate?}
\author{R Hasson}
\address{Department of Applied Mathematics, The Open University,
Milton Keynes, United Kingdom}
\begin{abstract}
In an inverse problem, such as the determination of brain activity
given magnetic field measurements outside the head, the main
quantity of interest is often the power associated with a source.
The `standard' way to determine this has been to find the best
linear estimate of the source and calculate the power associated
with this. This paper proposes an alternative method and then
relationship to this previous method of estimation is explored
both algebraically and by numerical simulation.

In abstract terms the problem can be stated as follows. Let $H$ be
a Hilbert space with inner product $\langle\ ,\ \rangle$. Let $L$
be a linear map: $H\rightarrow \R^n$. Suppose that we are given
data $b \in \R^n$ such that $b=Lx+e$ where $e$ is a vector of
random variables with zero mean and given covariance matrix which
represents measurement errors. The problem that is addressed in
this paper is to estimate $\langle x,\widehat{X}x \rangle$ where
$\widehat{X}$ is an operator on $H$ (e.g. the characteristic
function of a region of interest).

\end{abstract}
 \submitted
 \maketitle
\section{Introduction}

This paper solves a problem that arose in the study of the inverse
problem in magnetoencephalography (MEG)
\cite{Sarvas87,e:Hamalainen93}. The dominant concern in MEG
analysis has been to produce source maps of current density in the
brain and to co-register these to anatomical data (e.g.
\cite{Sarvas87,e:Schwartz96}). However, this may not be the most
appropriate approach when there is a focus on specific source
regions in the brain, e.g. the thalamus, fusiform gyrus etc. In
these cases it may be more appropriate to generate an activation
curve, a graph of the power dissipated in a specified region as a
function of time. Several methods of generating activation curves
have been proposed (e.g. \cite{p:Singh91,e:Tesche95,Robinson92}).
This aim of this paper is to derive an algorithm for generating
activation curves that is optimal with respect to the $L_2$-norm.

Another argument for the use of activation curves is the direct
comparison with other functional brain imaging modalities such as
positron emission tomography (PET) and functional magnetic
resonance imaging (fMRI). These modalities produce images of
quantities, e.g. regional cerebral blood flow (rCBF), that are
correlated with power dissipated rather than current density. This
suggests that in order to compare results across modalities we
should use magnetic field data to produce an estimate of the power
dissipated, i.e. an activation curve.

In Section~2 a more general problem is solved in the setting of a
linear map from a Hilbert space to a finite dimensional Hilbert
space. The main result from Section~2 (i.e.
Equation~\ref{eqn:final}) can be applied independently to each
time instant of the data from a MEG experiment. The method
proposed is to find a matrix $Y$ such that $b^TYb$ approximates
$\langle x,\widehat{X}x \rangle$ ($^T$ denotes matrix
transposition). The derivation of the optimal matrix $Y$
(Equation~\ref{eqn:final}) with respect to the $L_2$-norm is
contained in Section~2. Section~3 goes on to compare the main
results of Section~2 with the na\"\i ve algorithm which first
computes an estimate, \xreg, using Tikhonov regularization and
then computes $\langle \xreg ,\widehat{X}\xreg \rangle$. This
algorithm was used in \cite{p:Singh91} to extract measures of
brain activity.

In Section~4 we specialize to the study of the MEG inverse
problem. Definitions appropriate to this application are
introduced and a simulation study is described. In Section~5 an
important special case is considered where the region of interest
is the whole brain. A simplified equation
(Equation~\ref{eqn:simplifiedPower}) for this case is derived and
this is compared with the total signal power which is commonly
used as an estimate of brain activity. Section~6 is a discussion
of the merits of the algorithm together with the issues to be
addressed before applying the method in practice.

\section{Methods}
Let $H$ be a Hilbert space with inner product $\langle\ ,\
\rangle$. Let $L$ be a linear map: $H\rightarrow \R^n$. Suppose
that we are given data $b \in \R^n$ such that
\begin{equation}
b=Lx+e
\end{equation}
where $e$ is an unknown vector of random variables with zero mean
and covariance matrix $C$ which represents measurement error.
Suppose that the problem of finding an $x\in H$ corresponding to a
$b \in \R^n$ is an ill-posed problem. The problem here is to
estimate $\langle x,\widehat{X}x \rangle$ where $\widehat{X}$ is
an operator on $H$.

It should be noted that no assumptions are made about the noise in
the measurement channels other than it has zero mean and a well
defined covariance matrix $C$, i.e. if the measurement noise is
denoted by a vector $e$ then the covariance matrix is defined by
$C_{ij}=\overline{e_i e_j}$ where $\overline{\phantom{e_i}}$
denotes an expectation value.

Now define the adjoint map $L\adj$ by
\begin{equation}
\langle  x, L\adj b\rangle = (L x)^Tb, \qquad \mbox{for all } x\in
H, b \in \R^n.
\end{equation}
Here we are concerned with the image space $\cal I$ of $L\adj$.
Let $\{\widehat{e}_i: i=1\ldots n\}$ be the usual basis of $\R^n$
and choose a corresponding basis of $\cal I$, $\{\psi_i: i=1\ldots
n\}$, where $\psi_i=L\adj\widehat{e}_i$.

The matrix $Y$ will be chosen to minimize the error for points in
$\cal I$. The starting point in choosing an optimal matrix $Y$ is
to derive a suitable cost function to be minimized. We start by
expanding $b^TYb$.
\begin{equation} \label{eqn:start}
b^TY b= (Lx+e)^T Y (Lx+e) =(L x)^T Y L x+ e^T Y L x +(L x)^T Y e
+e^T Y e
\end{equation}
As mentioned above we focus on points in ${\cal I}\subseteq H$, so
we express $x\in \cal I$ in terms of our basis: $ x=\sum_{i=1}^n
a_i \psi_i$, where $a_i\in \R$ are scalars which will be written
collectively as a vector $a$. Equation~\ref{eqn:start} can be
simplified because the expression $Lx$ appears repeatedly, so
start by simplifying this expression:
\begin{equation}
  \label{eqn:Lx}
 (Lx)^T\widehat{e}_j = \langle x,L\adj\widehat{e}_j \rangle =
 \langle \Big( \sum_{i=1}^n a_i
 \psi_i\Big),\psi_j\rangle
  = \sum_{i=1}^n a_i \langle
\psi_i, \psi_j \rangle .
\end{equation}
The right hand side of Equation~\ref{eqn:Lx} can be written as the
$j$th component of a product $Pa$ where $P_{ij}=\langle \psi_i,
\psi_j \rangle$. Note that $P$ is a symmetric positive definite
$n\times n$ matrix. Substituting for $Lx$ in
Equation~\ref{eqn:start} gives:
\begin{equation} \label{eqn:bYbexpanded}
b^T Y b = a^T P Y P a+ e^T Y P a +a^T P Y e + e^T Y e.
\end{equation}
The projection of the operator $\widehat{X}$ onto $\cal I$ has a
matrix representation with respect to the basis $\{ \psi_i \}$
defined by $X_{ij} = \langle \psi_i, \widehat{X} \psi_j \rangle$
where $i,j=1,\ldots , n$. Hence the target expression can be
written in terms of the vector $a$:
\begin{equation}
\label{eqn:aXa} \langle x, \widehat{X} x \rangle = a^T X a,
\qquad \mbox{where}\ x=\sum_{i=1}^n a_i \psi_i .
\end{equation}
For $Y$ to be a good estimator, the right hand sides of
Equations~\ref{eqn:bYbexpanded} and \ref{eqn:aXa} should be
`close' for all $a\in \R^n$. One way of achieving this is to
minimize the cost function $E$ defined by:
\begin{equation} \label{eqn:error}
 E = \|X-PYP\|_2^2 + \|e^T YP\|_2^2 + \|P Y e\|_2^2 + \|e^T
 Ye\|_2^2.
\end{equation}
where $\|\ \|_2$ is the $L_2$-norm. Equation~\ref{eqn:error} can
be interpreted in physical terms. The first term is the error in
approximating the operator $\widehat{X}$ by $Y$. The second and
third terms give a measure of the overlap,induced by $Y$, between
the measurement error and the imaging space, $\cal I$. Note that
these terms are equal for a symmetric $Y$. The fourth term is a
measure of how $Y$ magnifies the measurement error.

To minimize $E$, $\partial E/\partial Y_{ik}$ is derived for each
element of the matrix $Y$. This gives $N^2$ equations to solve for
the $N^2$ unknowns $Y_{ik}$. These may be written as a single
matrix equation. In order to illustrate the manipulations
involved, the method will be elaborated for the fourth term in
Equation~\ref{eqn:error}. The fourth term is expanded using the
definition of the $L_2$-norm:
\begin{equation}
\|e^T Ye\|_2^2= \bigg( \sum_{\alpha ,\beta }e_{\alpha}
Y_{\alpha\beta} e_{\beta} \bigg) ^2.
\end{equation}
This is differentiated to obtain:
\begin{equation}
\frac{\partial \|e^T Ye\|_2^2}{\partial Y_{ik}} =
 2\bigg( \sum_{\alpha ,\beta }e_{\alpha} Y_{\alpha\beta} e_{\beta} \bigg)
e_i e_k = 2\sum_{\alpha ,\beta }e_i e_{\alpha} Y_{\alpha\beta}
e_{\beta}e_k .
\end{equation}
We proceed by replacing the products of random variables with
their expectation values, i.e.
 $\overline{e_i e_{\alpha}}=C_{i\alpha}$ and
 $\overline{e_{\beta} e_k}= C_{\beta k}$:
\begin{equation}
\frac{\partial \|e^T Y e\|_2^2}{\partial Y_{ik}} = 2\sum_{\alpha
,\beta }C_{i\alpha} Y_{\alpha\beta} C_{\beta k} .
\end{equation}
This is the $i k$th term of the matrix product $C Y\mskip -3mu C$.
Similarly, all of the other terms in Equation~\ref{eqn:error},
when differentiated, give terms that can be written as the $i k$th
elements of a product. So, the equations can be collected as:
\begin{equation}
-2PXP+2P^2 Y P^2 +2P^2 Y C+ 2CYP^2+ 2CYC=0.
\end{equation}
This may be written in the form:
\begin{equation}
(P^2 +C)Y(P^2 +C)= P X P.
\end{equation}
This equation can be solved in many ways, for example by defining
$Z=Y( P^2+C)$ and solving for $Z$ first and then for $Y$. This
easily implemented procedure was rejected as it computes an
non-symmetric $Y$ when starting with a symmetric matrix $X$,
because of the numerical problems associated with ill-conditioned
matrices. So an alternative scheme which preserves symmetry was
devised. Let $\lambda_i$ be the eigenvalue of the matrix $P$ with
eigenvector $\phi_i$. Then the matrices $X$ and $C$ can be
represented with respect to the basis $\{\phi_i\}$ as new matrices
$\Xnew $ and $\Cnew $, i.e.
\begin{eqnarray}
X&=&\sum_{ik} \phi^{}_i \Xnew ^{}_{ik} \phi_k^T, \qquad
  \mbox{where $\Xnew _{ik}= \phi_i^T X \phi^{}_k$}, \\
  C& =& \sum_{ik} \phi^{}_i \Cnew ^{}_{ik} \phi_k^T, \qquad
  \mbox{where $\Cnew _{ik}= \phi_i^T C \phi^{}_k$}.
\end{eqnarray}
With these definitions, the matrix $Y$ can be finally expressed as:
\begin{eqnarray}
Y & =& (P^2+C)^{-1}P \bigg(\sum_{ik} \phi^{}_i \Xnew ^{}_{ik}
\phi_k^T \bigg)P(P^2+C)^{-1}\\
 \label{eqn:final}
  & =&  \sum_{ik} \lambda^{}_i\lambda^{}_k\phi^{}_i (\Cnew +\lambda_i^2I)^{-1}
\Xnew ^{}_{ik} (\Cnew +\lambda_k^2I)^{-1} \phi_k^T
\end{eqnarray}

The matrix $Y$ computed using the above formula is always
symmetric for a given input symmetric matrix $X$.

Frequently the covariance matrix $C$ is not known and the
assumption is made that the random variables $e_i$ are independent
Gaussian random variables with a variance $\zeta$ that is
considered to be a parameter of the method. With this assumption
$C=\zeta I$ and Equation~\ref{eqn:final} becomes:
\begin{equation}\label{eqn:finalSpecial}
Y  = \sum_{ik} \frac{\lambda_i}{\lambda_i^2+\zeta}
\frac{\lambda_k}{\lambda_k^2+\zeta} \phi^{}_i \Xnew ^{}_{ik}
\phi_k^T
\end{equation}

\section{Comparison with na\"\i ve method}

We now compare Equation~\ref{eqn:finalSpecial} with the
corresponding equation derived by the na\"\i ve method mentioned
in the introduction. The na\"\i ve method for computing $\langle
x,\widehat{X}x \rangle$ is to compute a minimum norm estimate
using Tikhonov regularization to get $\xreg$ and then compute the
inner product.

To compute a $\xreg$ the first step is to choose a finite
dimensional subspace $R\subseteq H$ that has an orthonormal basis
$\{r_{\alpha}: \alpha=1,\ldots m\}$. The subspace $R$ will be
called the representation space and the regularized solution
$\xreg$ will lie in this space. The linear map $L:H\rightarrow
\R^n$ defines a linear map from $R$ to $\R^n$ by restriction that
we will also call $L$.

Now compute a singular value decomposition of $L:R \rightarrow
\R^n$ as $L=U\Sigma V^T$, where $\Sigma$ is a diagonal matrix with
non-negative entries $\sigma_1, \sigma_2, \ldots \sigma_n$ and $U$
and $V$ are matrices with orthonormal columns, i.e. $U^TU=V^TV=I$.
Applying Tikhonov regularization \cite{e:Hansen94} to the inverse
problem gives $\xreg= VDU^Tb$, where $D$ is a diagonal matrix
given by $D=(\Sigma^2+\zeta I)^{-1}\Sigma$. So the power
dissipated by this source can be computed by:
\begin{equation}
  \label{eqn:SVDcompareStep}
  \langle \xreg ,\widehat{X}\xreg \rangle =
  \left(b^TUDV^T\right){\cal X}\left(VDU^Tb\right),
\end{equation}
where $\cal X$ is the matrix representation of the operator
$\widehat{X}$ on $R$, i.e.
 ${\cal X}_{\alpha \beta}= \langle r_{\alpha}, \widehat{X} r_{\beta} \rangle$.
The right hand side of Equation~\ref{eqn:SVDcompareStep} is of the
form $b^T\widetilde{Y}b$ where $\widetilde{Y}$ is defined to be:
\begin{equation} \label{eqn:naiveY}
  \widetilde{Y} = UDV^T{\cal X}VDU^T
\end{equation}

The comparison with the method in the previous section relies on
the relationship between the linear operator $L$ and the
Gram-Schmidt matrix $P$ that we will now derive. Suppose for a
moment that the representation space $R$ was the whole of $H$ and
that the basis $\{r_{\alpha}\}$ is a complete orthonormal basis
for $R=H$. In this case:
\begin{eqnarray}
  P_{ij} = \langle \psi_i ,\psi_j \rangle &=&
  \sum_{\alpha}\langle \psi_i ,r_{\alpha} \rangle \langle r_{\alpha} ,\psi_j
  \rangle, \qquad \qquad \! \mbox{by completeness,}\\
  &=& \sum_{\alpha}\langle L\adj\widehat{e}_i ,r_{\alpha} \rangle \langle r_{\alpha}
  ,L\adj\widehat{e}_j \rangle, \qquad \mbox{using the definition of $\psi_i$,}\\
  &=&  \sum_{\alpha} \widehat{e}_i^T (L r_{\alpha}) (L r_{\alpha})^T
  \widehat{e}_j,
    \qquad \ \ \mbox{using the definition of $L\adj$,}\\
  &=&  \widehat{e}_i^T L
   \bigg(
     \sum_{\alpha}r^{}_{\alpha}r_{\alpha}^T
   \bigg)    L^T \widehat{e}_j, \qquad \: \mbox{by linearity,}\\
   &=& \widehat{e}_j^T LL^T \widehat{e}_i.
\end{eqnarray}
The right hand side of this equation is the $ij$th component of
the matrix product $LL^T$. So under the assumption that
$\{r_{\alpha}\}$ is a complete orthonormal basis for $H$ then
$P=LL^T$.

Now returning to the case when $R\subset H$ we can see that for a
good choice of $R$ the matrix $\widetilde{P}$ defined to be $LL^T$
will be approximately  equal to $P$. This is not surprising since
to compute the Gram-Schmidt matrix $P$ on a computer one usually
takes a suitable representation space $R$ and computes $LL^T$. The
singular value decomposition of $L$ immediately gives an
eigenvalue decomposition of $\widetilde{P}$ since
\begin{equation}
  \widetilde{P} = L L^T = U\Sigma V^TV \Sigma U^T = U \Sigma^2
  U^T,
\end{equation}
where the last equality follows from the fact that the columns of
$V$ are orthonormal. So the matrix $\widetilde{P}$ has eigenvalues
$\sigma_i^2$ with eigenvectors, $\widetilde{\phi}_i$ given by the
columns of $U$.

\begin{figure}[ht]\begin{center}
\begin{picture}(70,55)
\put(0,-5){
\put(10,10){\mbox{\epsfxsize=5cm\epsfysize=5cm\epsffile{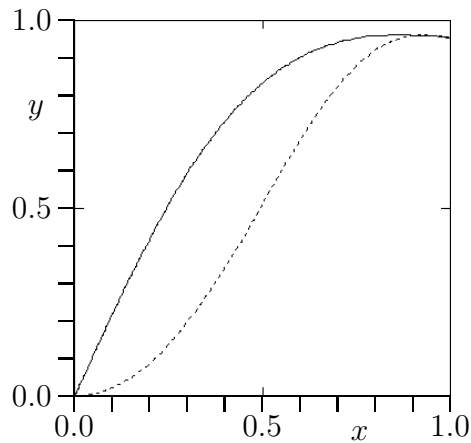}}
} \put(10,10){\line(1,0){50}} \put(10,10){\line(0,1){50}}
 \multiput(10,10)(5,0){11}{\line(0,-1){2}}
 \multiput(10,10)(0,5){11}{\line(-1,0){2}}
 \put(10,6){\makebox(0,0){0.0}}
 \put(35,6){\makebox(0,0){0.5}}
 \put(60,6){\makebox(0,0){1.0}}
 \put(48,6){\makebox(0,0)[t]{$x$}}
 \put(7,10){\makebox(0,0)[r]{0.0}}
 \put(7,35){\makebox(0,0)[r]{0.5}}
 \put(7,60){\makebox(0,0)[r]{1.0}}
 \put(6,48){\makebox(0,0)[r]{$y$}}
} %end of lh picture
\end{picture}
\end{center}
\caption{\label{fig:graphs} Graphs of the functions
$x/(x^2+\zeta)$ (solid curve) and $x^2/(x^4+\zeta)$ (dashed curve)
for $\zeta=0.5$. }
\end{figure}

By a similar argument to the above it can be seen that the matrix
$\widetilde{X}'$ defined to be $V^T{\cal X}V$ approximates the
matrix $\Xnew $ so we have:
\begin{equation}
  \label{eqn:naiveYspecial}
  \widetilde{Y}= U D \widetilde{\cal X} D U^T
  = \sum_{ik} \frac{\sigma_i}{\sigma_i^2+\zeta}
    \frac{\sigma_k}{\sigma_k^2+\zeta}
      \widetilde{\phi}^{}_i \widetilde{X}'^{}_{ik}
      \widetilde{\phi}_k^T,
\end{equation}
Now we can compare Equation~\ref{eqn:naiveYspecial} with
Equation~\ref{eqn:finalSpecial}. For a good representation space
$R$ we have $\widetilde{\phi}_i \simeq \phi_i$,
$\widetilde{X}'\simeq \Xnew $ and so the major difference between
the two approaches is that $\lambda_i \simeq \sigma_i^2$. The
effect of this change can be seen by plotting out the graphs of
the functions on the interval $[0,1]$ (this is the only range of
interest since we could dividing by the largest singular value
restrict to this interval). These graphs are shown in
Figure~\ref{fig:graphs} where it can be seen that
Equation~\ref{eqn:finalSpecial} attenuates the contribution from
the small singular values and has a sharper cut-off than is the
case for Equation~\ref{eqn:naiveYspecial}. The effect of this is
that Equation~\ref{eqn:finalSpecial} should attenuate the noise
component, which is usually associated with the small singular
values.
\section{Application}
Now we apply our results to the MEG inverse problem, i.e. the
problem of recovering information about source current density
inside the brain given measurements of the magnetic field outside
the brain. Let $\Omega$ denote the brain volume. The Hilbert space
of interest to us is, $L_2(\Omega )$, the space of square
integrable vector fields defined on the brain volume $\Omega$
together with the inner product:
\begin{equation}
\langle \vec{j}_1 ,\vec{j}_2 \rangle = \int_{\Omega}
\frac{\vec{j}_1(\vec{r})\cdot\vec{j}_2(\vec{r})}{\omega(\vec{r})}
{\rm d}\vec{r}, \quad \mbox{for all } \vec{j}_1 ,\vec{j}_2 \in
L_2(\Omega).
\end{equation}
The factor $\omega(\vec{r})$ is a weighting factor that allows
some flexibility in the procedure. The only restriction imposed on
$\omega(\vec{r})$ is that the integral over each voxel is finite.
In other papers the factor $\omega(\vec{r})$ has been interpreted
as a probability weight \cite{p:Hasson95D}.

It is interesting in this context to look at the the spatial
selectivity implicit in the use of the matrix $Y$ as it varies in
source space. Then the sensitivity profile of $Y$ at a point in
source space, $\vec{r}_0$, is defined to be
\begin{equation}
I(\vec{r}_0) = \sum_{i=1}^3 (L\vec{d}_{\vec{r}_0}^{\,i})^T Y
(L\vec{d}_{\vec{r}_0}^{\,i}),
\end{equation}
where $\vec{d}_{\vec{r}_0}^{\,i}$ is the current dipole
distribution, i.e.
$\vec{d}_{\vec{r}_0}^{\,i}(\vec{r})=\delta(\vec{r}-\vec{r}_0)\widehat{e}_i$
where $\{ \widehat{e}_i : i=1,2,3\}$ is an orthogonal set of unit
vectors and $\delta(\ )$ denotes the Dirac delta function.

The spatial selectivity, $I(\vec{r}_0)$, may be thought of as an
instrumental generalization of the lead field of a single
measurement channel. The definition is designed so that in the
case when $Y_{ik}=1$ when $i=k=n_0$ and 0 otherwise then the
sensitivity $I(\vec{r}_0)$ is the square of the magnitude of the
lead field of channel $n_0$. Note that the above definition of
$I(\vec{r}_0)$ is different from the original definition proposed
in \cite{p:Hasson98A}.

\begin{figure}[ht]
\begin{center}
\begin{picture}(130,55)
\put(0,-5){
\put(10,10){\mbox{\epsfxsize=5cm\epsfysize=5cm\epsffile{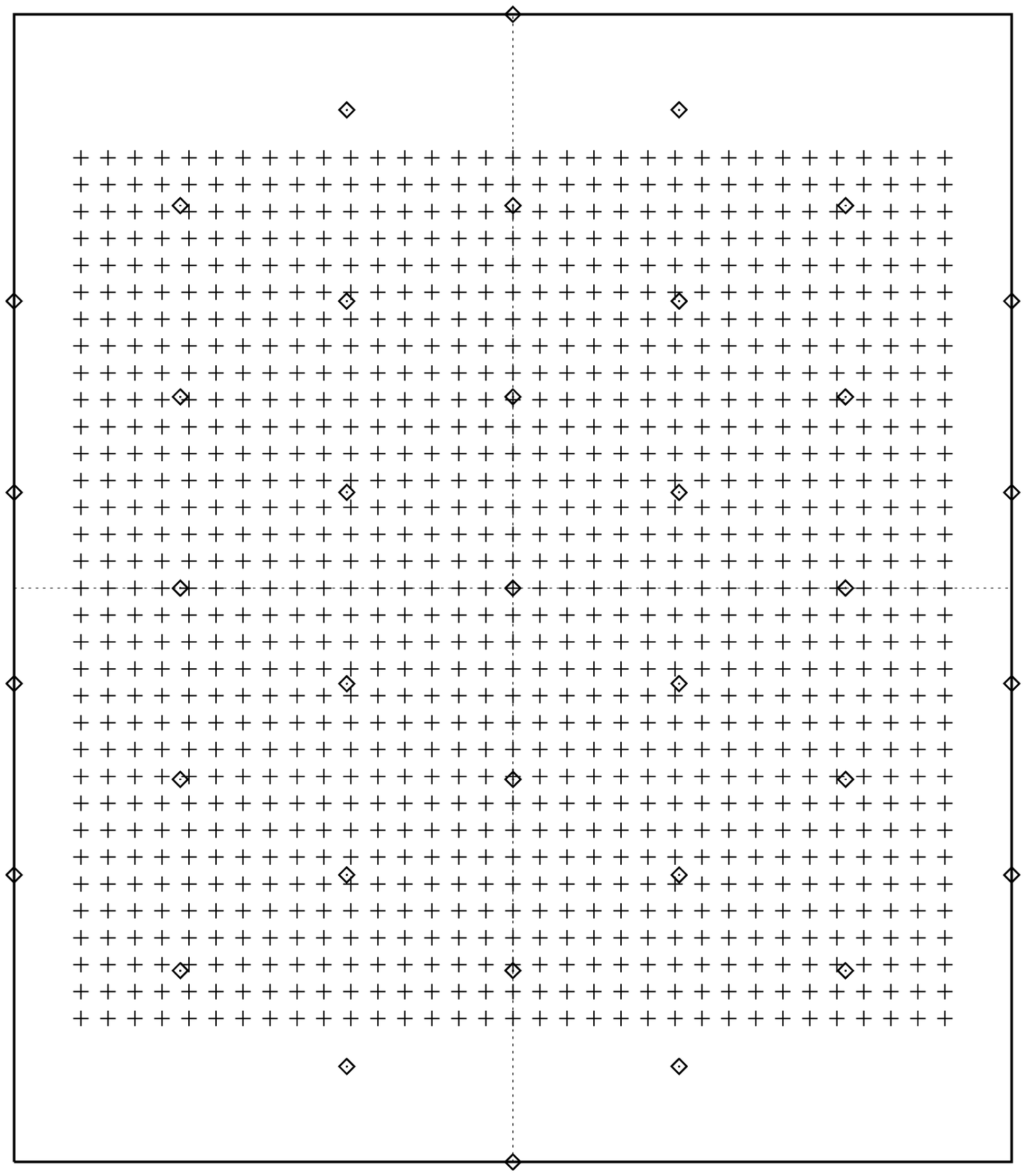}}
} \put(10,10){\line(1,0){50}} \put(10,10){\line(0,1){50}}
\multiput(10,10)(5,0){11}{\line(0,-1){2}}
\multiput(10,10)(0,5){11}{\line(-1,0){2}}
\put(10,6){\makebox(0,0){-0.06}} \put(35,6){\makebox(0,0){0.0}}
\put(60,6){\makebox(0,0){0.06}}
\put(48,6){\makebox(0,0)[t]{$x$/m}}
\put(7,10){\makebox(0,0)[r]{-0.06}}
\put(7,35){\makebox(0,0)[r]{0.0}}
\put(7,60){\makebox(0,0)[r]{0.06}}
\put(6,48){\makebox(0,0)[r]{$y$/m}}
} %end of lh picture
\put(65,-5){
\put(10,10){\mbox{\epsfxsize=5cm\epsfysize=5cm\epsffile{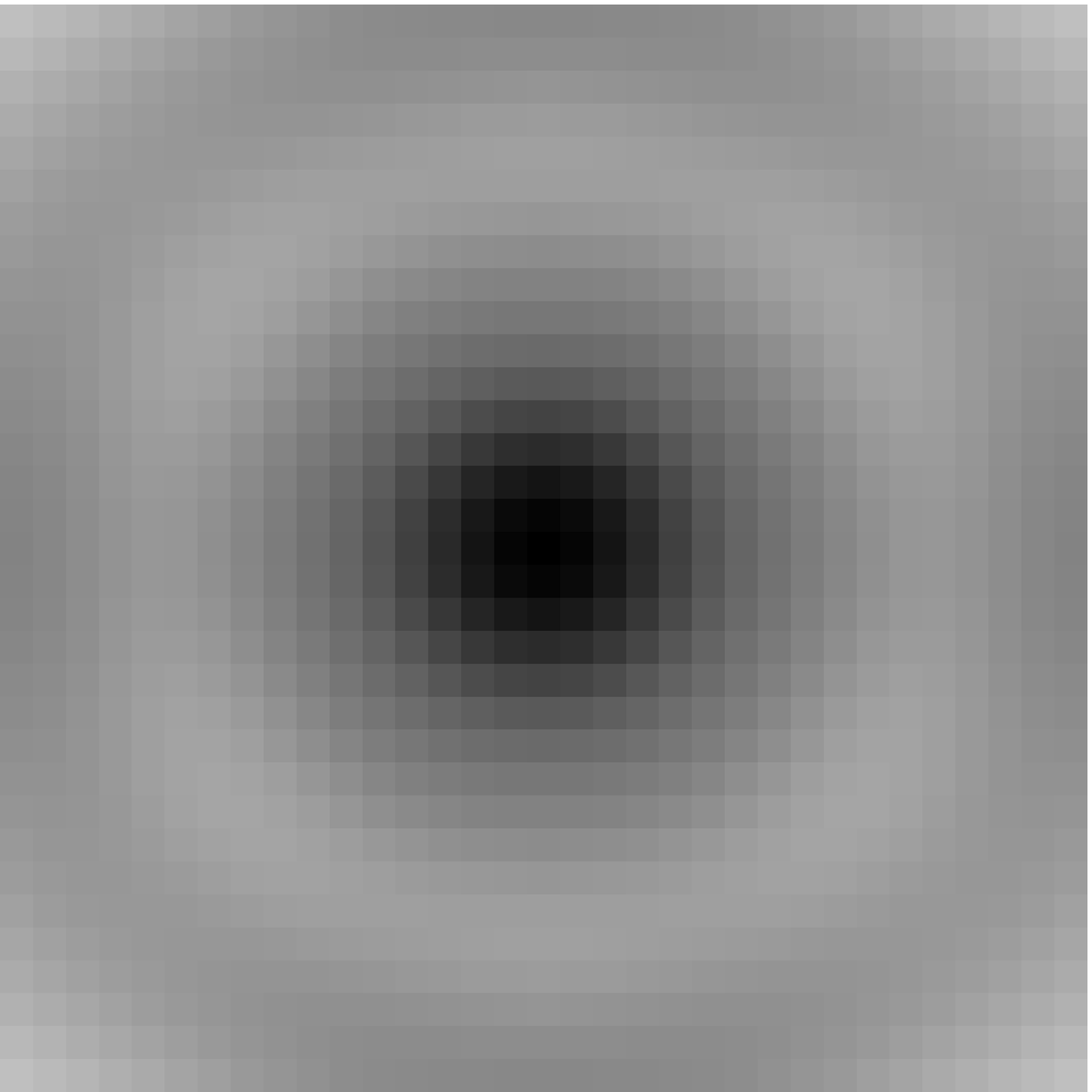}}
} \put(10,10){\line(1,0){50}} \put(10,10){\line(0,1){50}}
\multiput(10,10)(5,0){11}{\line(0,-1){2}}
\multiput(10,10)(0,5){11}{\line(-1,0){2}}
\put(10,6){\makebox(0,0){-0.045}} \put(35,6){\makebox(0,0){0.0}}
\put(60,6){\makebox(0,0){0.045}}
\put(48,6){\makebox(0,0)[t]{$x$/m}}
\put(7,10){\makebox(0,0)[r]{-0.045}}
\put(7,35){\makebox(0,0)[r]{0.0}}
\put(7,60){\makebox(0,0)[r]{0.045}}
\put(6,48){\makebox(0,0)[r]{$y$/m}}
} %end of rh picture
\end{picture}
\end{center}
\caption{\label{fig:geometry} (left) A plan view of the experiment
geometry. Crosses denote source space voxels and diamonds denote
the projections of the centres of the detector coils. (right) The
sensitivity profile in source space of the $Y$ matrix that is
derived from the operator $\widehat{X}=\delta(\vec{r}-\vec{r_c})$.
}
\end{figure}

To illustrate the method a simple simulated experimental system
(Figure~\ref{fig:geometry}) has been investigated. The head is
modelled as a homogeneous conducting sphere of radius $8.9$\,cm
with its centre at ($0,0,-0.07\,$cm). The source space is a
$9$\,cm$\times9$\,cm square thin lamina consisting of $33\times33$
voxels in the plane $z=-0.01$\,cm with centre ($0,0,-0.01\,$cm).
The measurement instrument is a hexagonal array of $37$ second
order axial gradiometers with baseline $5$\,cm with the lowest
'sensing' coils in the plane $z=4$\,cm.
Now consider, in the context of the simulated system, the simplest
possible region of interest operator
$\widehat{X}=\delta(\vec{r}-\vec{r}_c)$ where
$\vec{r}_c=(0,0,-0.01$\,cm) is the centre of source space. This
type of operator might be adopted if one simply wished to focus on
a small volume of source space. The matrix $Y$ used as an
estimator from this operator is calculated using
Equation~\ref{eqn:finalSpecial}. The sensitivity profile for this
$Y$ matrix is shown in Figure~\ref{fig:geometry}.

The reconstruction of an activation curve has been tested on
simulated data using this region of interest operator and
simulated data from a time varying target dipole at ($0,0,0\,$cm),
i.e. 1\,cm from the region of interest. The moment of the dipole
varies sinusoidally at 10\,Hz, with an envelope that rises
linearly from zero at 200\,ms to a maximum at 300\,ms after which
it remains constant. To show the insensitivity to dipole
orientation the dipole moment was made to rotate smoothly in a
tangential plane --- this rotation is not discernible in the
activation curve. In addition to the target dipole there is
distractor dipole at ($0,\,0.02\,\mbox{cm},\,0$), which is active
from 0 to 100\,ms (triangular envelope) and again from 400\,ms
(square envelope).

In the period from 200\,ms to 400\,ms when only the target dipole
is active, the calculated (power) activation curve matches closely
that of the target. However, the existence of the distractor
dipole within the sensitive region (see Figure~\ref{fig:geometry})
gives rise to apparent activity between 0\,ms and 100\,ms and
inaccuracy in the calculated activation curve for the period after
400\,ms. The distractor dipole adds to the estimated power
dissipated when it is parallel to the target and subtracts when
the target dipole has rotated to be anti-parallel.

\begin{figure}[ht]
\begin{center}
\begin{picture}(110,65)
 \put(10,5){\mbox{\epsfxsize=10cm\epsfysize=6cm\epsffile{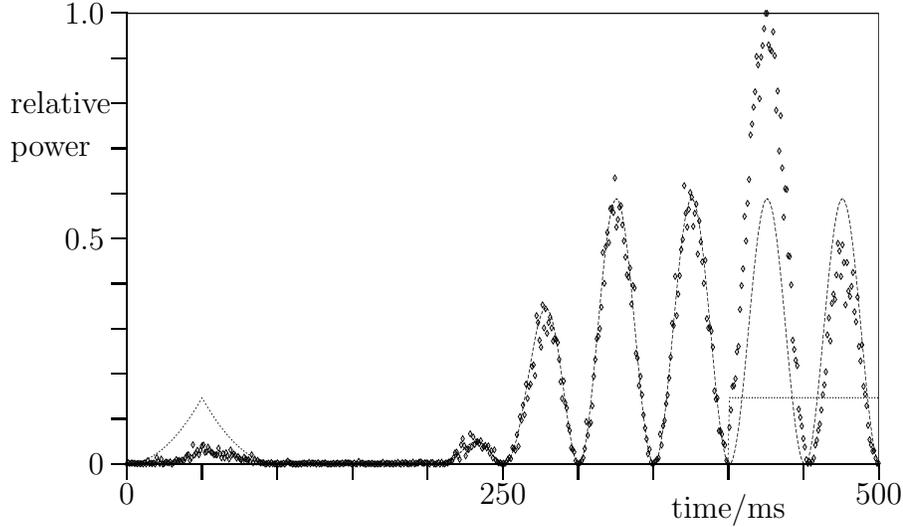}}}
 \put(10,5){\line(1,0){100}}
 \put(10,5){\line(0,1){60}}
 \multiput(10,5)(10,0){11}{\line(0,-1){2}}
 \multiput(10,5)(0,6){11}{\line(-1,0){2}}
 \put(10,1){\makebox(0,0){0}}
 \put(60,1){\makebox(0,0){250}}
 \put(110,1){\makebox(0,0){500}}
 \put(90,1){\makebox(0,0)[t]{time/ms}}
 \put(7,5){\makebox(0,0)[r]{0}}
 \put(7,35){\makebox(0,0)[r]{0.5}}
 \put(7,65){\makebox(0,0)[r]{1.0}}
 \put(10,50){\makebox(0,0)[r]{\begin{tabular}{l}relative\\power\end{tabular}}}
\end{picture}
\end{center}
\caption{\label{fig:act} Activation curves for a simulated
experiment. The solid line and the dotted lines are the activation
curves of the target and distractor dipoles. The diamonds are the
calculated activation curve from the $Y$ matrix whose sensitivity
profile is shown in Figure~\ref{fig:geometry}. The error bars,
omitted for clarity, would be approximately twice the height of
the diamonds.}
\end{figure}
Error bars for the activation curve can be estimated using the
last term in Equation~\ref{eqn:error} to give the amount of
measurement noise reflected in the activation curve. The estimate
is given by $\sum_{\alpha, \beta} C_{\alpha\beta}Y_{\alpha\beta}$.

\section{Total brain activity}
As a special case of Equation~\ref{eqn:finalSpecial} the task of
finding an estimate of the total activity in the source space is
considered. In this case the operator $\widehat{X}$ is the
identity and so
\begin{equation}
  X_{ij}= \langle  \psi_i, \widehat{X}\psi_j \rangle
  = \langle  \psi_i, \psi_j \rangle = P_{ij}
\end{equation}
So the matrix $\Xnew $ can be calculated as follows
\begin{equation}
 \Xnew _{ij} = \phi_i^T X \phi^{}_j = \phi_i^T P \phi^{}_j
 = \lambda^{}_j \phi_i^T \phi^{}_j = \lambda_j \delta _{ij}
\end{equation}
where $\delta _{ij}$ is the Kronecker delta. So, in this case, $Y$
is given by the simplified formula:
\begin{equation}
 Y = \sum_{ij} \frac{\lambda_i}{\lambda_i^2+\zeta}  \frac{\lambda_j}{\lambda_j^2+\zeta}
 \phi^{}_i  \lambda^{}_j \delta^{}_{ij} \phi_j^T
 = \sum_{i} \frac{\lambda_i^3}{(\lambda_i^2+\zeta)^2}\phi^{}_i \phi_i^T
\end{equation}
This gives the following formula for computing the total activity.
\begin{equation}
\label{eqn:simplifiedPower}
 \mbox{Total activity, }A(t) = \sum_{i}
    \frac{\lambda_i^3}{(\lambda_i^2+\zeta)^2}
    \left( \phi_i^T b(t)\right)^2
\end{equation}
where $b(t)$ is the vector of measurements collected at time $t$.

\begin{figure}[ht]
\begin{center}
\begin{picture}(130,35)
\put(10,5){
  \put(0,0){\mbox{\epsfxsize=5cm\epsfysize=3cm\epsffile{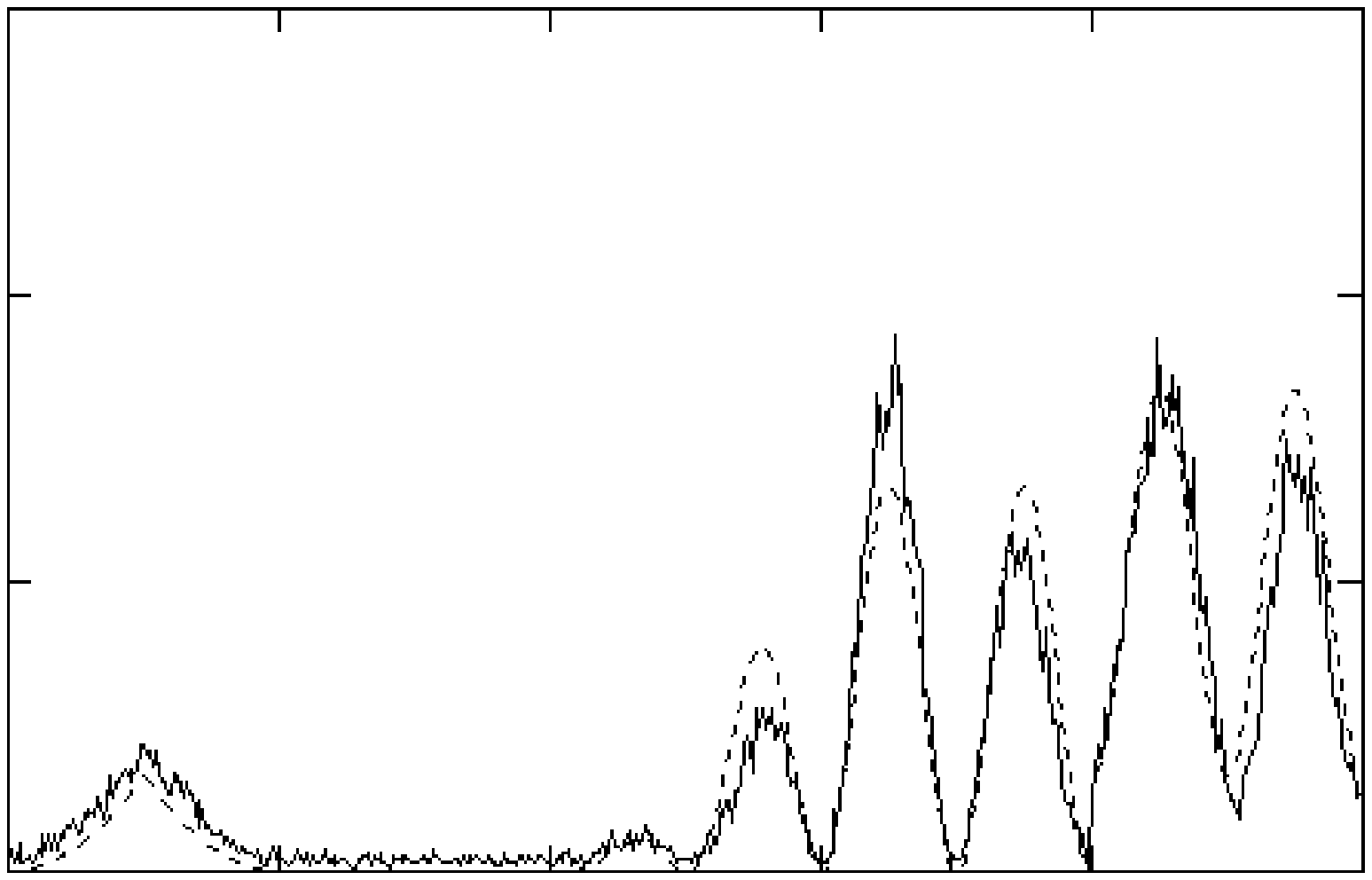}}}
  \put(0,0){\line(1,0){50}} %x axis
  \put(0,0){\line(0,1){30}} %y axis
%  \multiput(10,10)(10,0){11}{\line(0,-1){2}} %x ticks
%  \multiput(10,10)(0,6){11}{\line(-1,0){2}}  %y ticks
  \put( 0,-4){\makebox(0,0){0}}
  \put(25,-4){\makebox(0,0){250}}
  \put(50,-4){\makebox(0,0){500}}
  \put(37.5,-4){\makebox(0,0)[t]{time/ms}}
  \put(-1, 0){\makebox(0,0)[r]{0}}
  \put(-1,10){\makebox(0,0)[r]{0.005}}
  \put(-1,20){\makebox(0,0)[r]{0.01}}
  \put(-1,30){\makebox(0,0)[r]{0.015}}
  \put(-2,25){\makebox(0,0)[r]{$A(t)$}}
} %end of lh picture
\put(80,5){
  \put(0,0){\mbox{\epsfxsize=5cm\epsfysize=3cm\epsffile{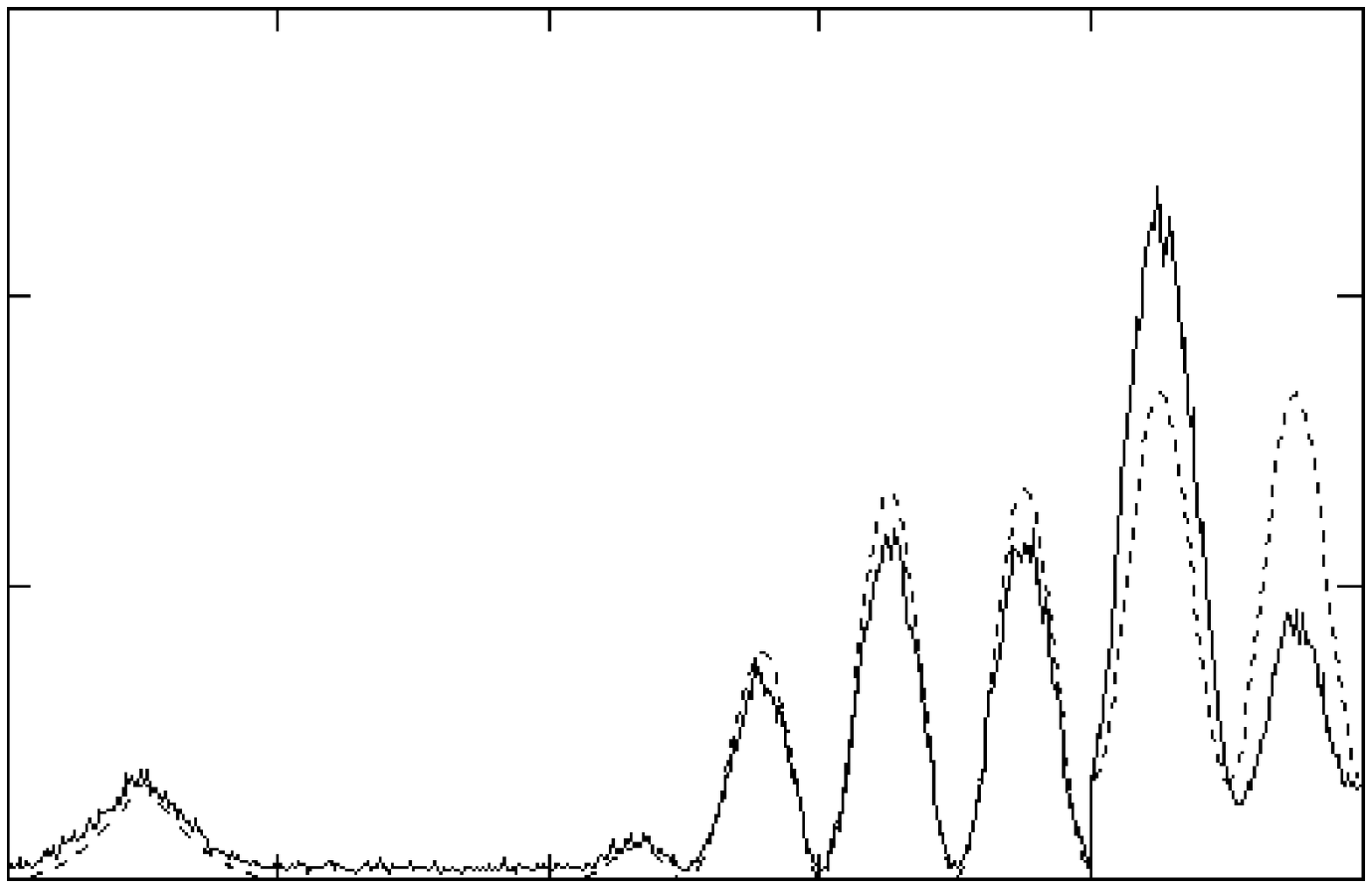}}}
  \put(0,0){\line(1,0){50}} %x axis
  \put(0,0){\line(0,1){30}} %y axis
%  \multiput(10,10)(10,0){11}{\line(0,-1){2}} %x ticks
%  \multiput(10,10)(0,6){11}{\line(-1,0){2}}  %y ticks
  \put( 0,-4){\makebox(0,0){0}}
  \put(25,-4){\makebox(0,0){250}}
  \put(50,-4){\makebox(0,0){500}}
  \put(37.5,-4){\makebox(0,0)[t]{time/ms}}
  \put(-1, 0){\makebox(0,0)[r]{0}}
  \put(-1,10){\makebox(0,0)[r]{0.005}}
  \put(-1,20){\makebox(0,0)[r]{0.01}}
  \put(-1,30){\makebox(0,0)[r]{0.015}}
  \put(-1,25){\makebox(0,0)[r]{$B(t)$}}
} %end of rh picture
\end{picture}
\end{center}
\caption{\label{fig:powernpower}(left) A comparison of the total
brain activity, $A(t)$, (solid line) with a plot of the power of
the dipolar sources that generated the simulated data (dashed
line). In order to compare with the right-hand diagram both curves
are normalized to enclose a unit area. (right) A comparison of the
total signal power, $B(t)$, (solid line) with a plot of the power
of the dipolar sources that generated the simulated data (dashed
line). In order to compare with the left-hand diagram both curves
are normalized to enclose a unit area.}
\end{figure}

Previously when an estimate of the total brain activity was needed
the power in the signals was used, i.e.
\begin{equation}
 \mbox{Total signal power, }B(t) = b(t)^T b(t)
\end{equation}
These two methods have been compared for the simulated data
described above as shown in Figure~\ref{fig:powernpower}. In
Figure~\ref{fig:powernpower} it can be seen that the estimate
$A(t)$ (shown as the solid line on the left) more closely
approximates the true activation of the dipoles (dashed curve)
than the estimate $B(t)$. In fact, if the error in the estimate is
measured by the integral of the squared discrepancies between the
curves then the error for $A(t)$ is $2.6\times 10^{-4}$ whilst the
error for $B(t)$ is $6.6\times 10^{-4}$.
\section{Discussion}
We have shown that it is possible to directly compute the `power'
associated with a source without computing the source first. The
method seems robust to noise and is not dependent on the noise
having a Gaussian profile. Correlations between measurement
channels are fully taken into account. In particular it was shown
that activation curves of brain regions can obtained from magnetic
field data. The method provides an easily computable way of
tracking the power dissipated in a specific region of the brain.

To use the method effectively the practical problem is to
effectively estimate the covariance matrix. For evoked response
experiments the covariance matrix, $C$, can be estimated from the
prestimulus period. For other experiments it might be more
suitable to make the {\em a priori} assumption that the noise is
uncorrelated Gaussian noise with variance a $\alpha^2$ that could
be considered as a parameter. As $\alpha$ increases, the more
closely the $Y$ matrix sensitivity pattern matches the region of
interest, but the larger the error bars on the resulting
activation curve.

Finally, to answer the question in the title, I would say that if
best is interpreted in a least $L_2$-norm sense then the answer is
no. The best way to estimate the power associated with a source is
to compute it directly.

\section*{References}

\bibliographystyle{unsrt}
\bibliography{biomag74,biomag79,biomag82,biomag84,biomag86,biomag88,biomag90,biomag91,biomag92,muenster,personal,extra}

\end{document}